\documentclass[aps,prl,twocolumn,superscriptaddress]{revtex4}
\usepackage{graphicx}
\usepackage{amssymb}
\usepackage{amsmath}
\usepackage{graphicx}
\usepackage{epsfig}
\usepackage{array}
\usepackage{color}
\usepackage{multirow}

\begin{document}

\title{Hund-driven local-itinerant duality of Eu-4$f$ electrons in infinite-layer nickelates}

\author{Yingying Cao}
\email[]{caoyingying@hnas.ac.cn}
\affiliation{Institute of Quantum Materials and Physics, Henan Academy of Sciences, Zhengzhou 450046, China}
\author{Yi-feng Yang}
\email[]{yifeng@iphy.ac.cn}
\affiliation{Beijing National Laboratory for Condensed Matter Physics and
Institute of Physics, Chinese Academy of Sciences, Beijing 100190, China}
\affiliation{University of Chinese Academy of Sciences, Beijing 100049, China}

\date{\today}

\begin{abstract}
Recent discovery of reentrant superconductivity and elevated $T_c$ in Eu-substituted infinite-layer nickelates raises a critical question concerning the nature of Eu-4$f$ states and their potential interaction with Ni-$d$ electrons. Here we combine density functional theory with the dynamical mean-field theory (DFT+DMFT) calculations to investigate the valence and magnetic properties of Eu-ions in the nickelate 112 structure, explicitly treating the Coulomb repulsion in both Ni-3$d$ and Eu-4$f$ shells. We find a three-regime evolution of the Eu-4$f$ configuration driven by its Hund's rule coupling $J_{\rm H}$. The magnetic and mixed-valence Eu ionic state inferred by experiments only occurs for moderate $J_{\rm H}$, with coexisting local moments and strongly correlated $j=7/2$ quasiparticles, in contrast to the nonmagnetic state at small $J_{\rm H}$ and the fully spin-polarized divalent state at large $J_{\rm H}$. This local-itinerant duality arises from an effectively hole-doped orbital-selective Mott state induced by the intra-4$f$ charge transfer between $j=5/2$ and $7/2$ manifolds. It not only explains the origin of the mixed-valence and local moment behaviors of the Eu-ions observed experimentally, but also predicts low-energy 4$f$ quasiparticles that may hybridize with the Ni-3$d$ electrons and contribute to the elevated $T_c$. Our work provides a basis for understanding the unusual ferromagnetic and superconducting properties of Eu-substituted infinite-layer nickelates.
\end{abstract}

\maketitle

Superconductivity in infinite-layer nickelates $A$NiO$_2$ ($A$ = La, Pr, Nd, Sr, Ca) has attracted intensive interest since its discovery as a new platform for investigating unconventional pairing mechanisms \cite{Li2019N, Li2020PRL, Zeng2022SA, Osada2020NL, Osada2020PRM, Wei2023SA,Parzyck2025PRX}. The Ni$^{1+}$ ions in these compounds have a $3d^9$ configuration analogous to that of Cu$^{2+}$ in cuprate high-$T_c$ superconductors. However, first-principles calculations reveal that the low-energy electronic structure involves not only the primary Ni-$d_{x^2-y^2}$ orbital but also O-$p$ orbitals, rare-earth 5$d$ and other Ni-$d$ orbitals, which together shape the overall Fermi surface topology \cite{Karp2020PRX, Jiang2019PRB,Hepting2020NM,Been2021PRX}. Two central issues emerge concerning the role of charge doping into Ni-$d_{x^2-y^2}$ orbitals and the nature of magnetic fluctuations responsible for the electron pairing \cite{Yang2022FP,Gu2022I,Zhang2020PRBa,Kitatani2020NQM,Cui2021CPL,Wu2020PRB,Wang2020PRBa}. One prevailing view treats the Ni-$d_{x^2-y^2}$ orbital as the single active channel for superconductivity, with the rare-earth ions (e.g., La$^{3+}$) acting as charge reservoirs that supply carriers through hybridization with O-$2p$ and non-$d_{x^2-y^2}$ Ni states. Within this single-band Hubbard model framework, the doping dependence of $T_{c}$ was well reproduced through a combined DFT+DMFT and dynamical vertex approximation approach \cite{Kitatani2020NQM}. Alternatively, the interstitial $s$-orbital has also been proposed to play a crucial role in the superconductivity and charge density order \cite{Wang2020PRBa,Gu2020CP,Chen2023NC}. Wannier analyses revealed substantial interstitial-$s$ character in the relevant low-energy bands \cite{Gu2020CP, Nomura2019PRB}. Scanning tunneling spectroscopy revealed multiple superconducting gap features \cite{Gu2020NC}, while the charge-density order has been linked to hybridized Ni-$3d$/Nd-$5d$ bands \cite{Tam2022NM}. Angle-resolved photoemission spectroscopy (ARPES) measurements on NdNiO$_2$ and LaNiO$_2$ further confirmed its contribution to the Fermi surface topology and its hybridization with the Ni-$d$ bands \cite{Sun2025SA,Li2025PRL}. Moreover, the correlation between $T_c$ and the $c$-axis lattice parameter suggests that interlayer coupling, beyond a simple single-band picture, is also relevant \cite{Yang2026NC}.

Recently, substitution of Sm or Eu for the A-site elements in the 112-type nickelate superconductors has drawn increasing attention, as these systems exhibit the enhanced $T_c$ at ambient pressure \cite{Chow2025N, Han2026NSR,Yang2026N}. X-ray absorption near-edge spectroscopy (XANES) measurements at the Eu $L_3$ edge on $(\mathrm{Nd},\mathrm{Eu})$NiO$_2$ reveal a mixed-valence state of Eu ions, suggesting a coupling between the Eu and Ni valences \cite{Wei2023SA}. Consistently, Eu $M$-edge X-ray absorption spectroscopy measurements in $(\mathrm{Sm,Eu,Ca})\mathrm{NiO}_2$ identified mixed Eu$^{2+}$/Eu$^{3+}$ states \cite{Yang2026N}. Moreover, reentrant superconductivity has been observed in several Eu-substituted infinite-layer nickelates. In $(\mathrm{Sm,Eu,Ca})\mathrm{NiO}_2$, field reentrant superconductivity was reported in the Eu-substituted regime \cite{Yang2026N}. Related studies also reported reentrant or field-enhanced superconductivity in $(\mathrm{Nd,Pr,Eu})\mathrm{NiO}_2$ and $(\mathrm{Nd,Dy,Eu})\mathrm{NiO}_2$ \cite{Han2026NSR,Varbaro2026NC,Vu2026NC,Yang2026}. Separately, low-temperature magnetic freezing and a substantial net magnetization have been reported in $(\mathrm{Sm,Eu,Ca})\mathrm{NiO}_2$ \cite{Grutter2025}. These reveal a dual role of the Eu ions, whose nature has not been clarified and warrants further theoretical and experimental scrutiny.

In this work, we employ the DFT+DMFT method to investigate the Eu ionic state in the ideal stoichiometric EuNiO$_2$, treating the Coulomb repulsion within both Ni-$3d$ and Eu-$4f$ shells. Our calculations reveal that the valence state of Eu ions depends sensitively on its Hund's rule coupling $J_{\rm H}$, leading to three distinct regimes. The magnetic and mixed-valence regime occurs only at moderate $J_{\rm H}$, where electrons are transferred progressively from $j=5/2$ to $j=7/2$ orbitals, causing fully polarized $j=5/2$ local moment and partially filled $j=7/2$ orbitals with coexisting local moment and strongly correlated quasiparticles near the Fermi energy. Such a dual character differs from the nonmagnetic state at small $J_{\rm H}$ and the fully polarized divalent state at large $J_{\rm H}$, suggesting that the Hund's rule coupling of $4f$ electrons plays a critical role in explaining the magnetic and mixed-valence properties of Eu ions observed in experiment. Our work clarifies the origin of the Eu spin and valence states in the 112 structure and thereby provides a potential basis for understanding the ferromagnetism and unusually high superconducting $T_c$ in Eu-substituted infinite-layer nickelates.

\begin{figure}[tb]
    \begin{center}
        \includegraphics[width=0.48\textwidth]{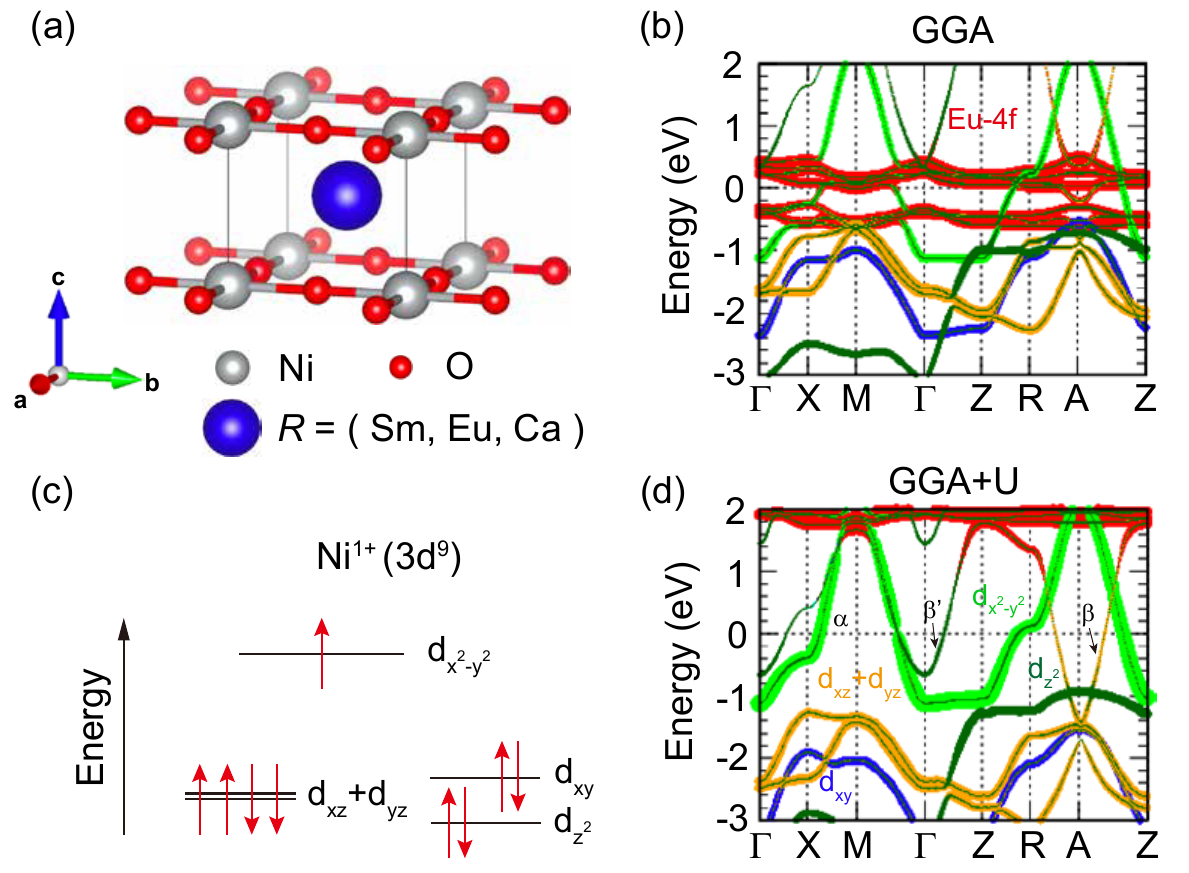}
        \caption{(a) Illustration of the crystal structure of infinite-layer nickelates with stacked NiO$_2$ layers and rare-earth elements in between. (b) Orbital-projected band structure from generalized-gradient approximation (GGA) calculations. (c) Schematic representation of the 3$d^9$ configuration for Ni$^{1+}$ ions. (d) Orbital-projected band structure from GGA+$U$ calculations, with onsite Coulomb repulsions for both Eu-$4f$ and Ni-$d$ orbitals.}\label{fig1}
    \end{center}
\end{figure}

A schematic plot of the 112 structure is given in Fig.~\ref{fig1}(a), where Ni ions occupy the vertices of a square lattice, with O ions at the edge midpoints forming the NiO$_2$ plane. These layers stack along the $c$-axis, with rare-earth (La, Nd, Pr, Sm, Eu) or alkaline-earth (Ca, Sr) ions intercalated in between above the square centers. To elucidate the ionic state of Eu, we focus on the ideal $A$-site single-element bulk compound EuNiO$_2$. First-principles calculations were performed using the full-potential linearized augmented plane-wave method as implemented in the WIEN2k package \cite{Blaha2014,Blaha2020JCP}, with the Perdew-Burke-Ernzerhof generalized-gradient approximation (GGA-PBE) for the exchange-correlation functional \cite{Perdew1996PRL} and the optimized lattice parameters $a=b=3.859$ \AA\, and $c=$ 3.190 \AA\,. The Muffin-tin radii were set to 2.3 a.u. for Eu, 1.77 a.u. for Ni, and 1.52 a.u. for O. The plane-wave cutoff was defined by $R_{\rm MT}K_{\rm max} = 8.08$, and a $28 \times 28 \times 33$ ${\bf k}$-point mesh was used for the Brillouin-zone integration. The GGA+$U$ calculations were carried out with $U^{\rm Ni} = 4$ eV, $J_{\rm H}^{\rm Ni} = 0.7$ eV for Ni-$3d$ orbitals and $U = 8$ eV, $J_{\rm H} = 1$ eV for Eu-$4f$ orbitals according to previous studies on infinite-layer nickelates \cite{Kitatani2020NQM,Nomura2019PRB} and Eu-involved compounds \cite{Riley2018NC,Lazewski2021IC}.

Figure~\ref{fig1}(b) shows the GGA band structure of EuNiO$_2$. The low-energy electronic states near $E_{\rm F}$ are dominated by Ni-3$d$ and Eu-4$f$ orbitals. The Ni-3$d$ states exhibit a pronounced crystal-field splitting, with the partially occupied $d_{x^2-y^2}$ orbital separated from the nearly fully occupied $d_{z^2}$, $d_{xy}$, and $d_{xz}+d_{yz}$ manifolds. The resulting orbital occupations are consistent with an overall Ni-3$d^9$ configuration, whose crystal-field-level scheme in the NiO$_2$ plane is schematically illustrated in Fig.~\ref{fig1}(c). The Eu ions show a nominally trivalent state with a fully occupied $j=5/2$ multiplet near $-0.5$ eV and a $j=7/2$ manifold lying mainly above $E_{\rm F}$. In GGA+$U$ [Fig.~\ref{fig1}(d)], the large onsite Coulomb repulsion pushes the $j=5/2$ orbitals down to approximately $-8$ eV and the $j=7/2$ orbitals to well above $E_{\rm F}$, further locking Eu into the Eu$^{3+}$ configuration and leaving the low-energy spectrum dominated by Ni-$3d$ orbitals. The $d_{x^2-y^2}$ band crosses $E_{\rm F}$ and defines the prominent $\alpha$ sheet. Two other Fermi pockets, including a $\beta'$ pocket around $\Gamma$ and a $\beta$ pocket around A, have been associated with the rare-earth 5$d$ and interstitial $s$ orbitals \cite{Gu2020CP}. Obviously, neither GGA nor GGA+$U$ captures the mixed-valence and local-moment physics of Eu ions in the 112 structure, which motivates us to perform the state-of-the-art DFT+DMFT calculations below.

The DFT+DMFT Hamiltonian was constructed from the Kohn-Sham bands within an energy window of $[-10, 10]$ eV relative to the Fermi energy, which encompasses all Ni-$3d$ and Eu-$4f$ orbitals. The continuous-time hybridization-expansion (CT-HYB) quantum Monte Carlo method was employed as the impurity solver \cite{Haule2007PRB}, with the interaction parameters $U^{\rm Ni}=4$ eV and $J_{\rm H}^{\rm Ni}=0.7$ eV fixed for the Ni-$3d$ shell, while those for the Eu-$4f$ shell were varied over the ranges $U=4$--$8$ eV and $J_{\rm H}=0$--$1$ eV according to typical Eu-4$f$ values \cite{Riley2018NC,Lazewski2021IC,Zeer2025NCM}. Each calculation accumulated more than $10^7$ Monte Carlo steps across eight processors. The self-energy was then analytically continued using the maximum entropy method \cite{Jarrell1996PR}. The nominal scheme was applied for the double-counting correction \cite{Haule2015PRL}, with the reference occupations $n_d=8.4$ and $n_f=6.6$ corresponding to an average Eu valence of +2.4 as inferred from XANES measurements on $(\mathrm{Nd},\mathrm{Eu})$NiO$_2$ \cite{Wei2023SA}. Other double-counting schemes were also tested, but failed to yield a mixed-valence state of the Eu ions consistent with experimental observations.

\begin{figure}[tb]
    \begin{center}
        \includegraphics[width=0.48\textwidth]{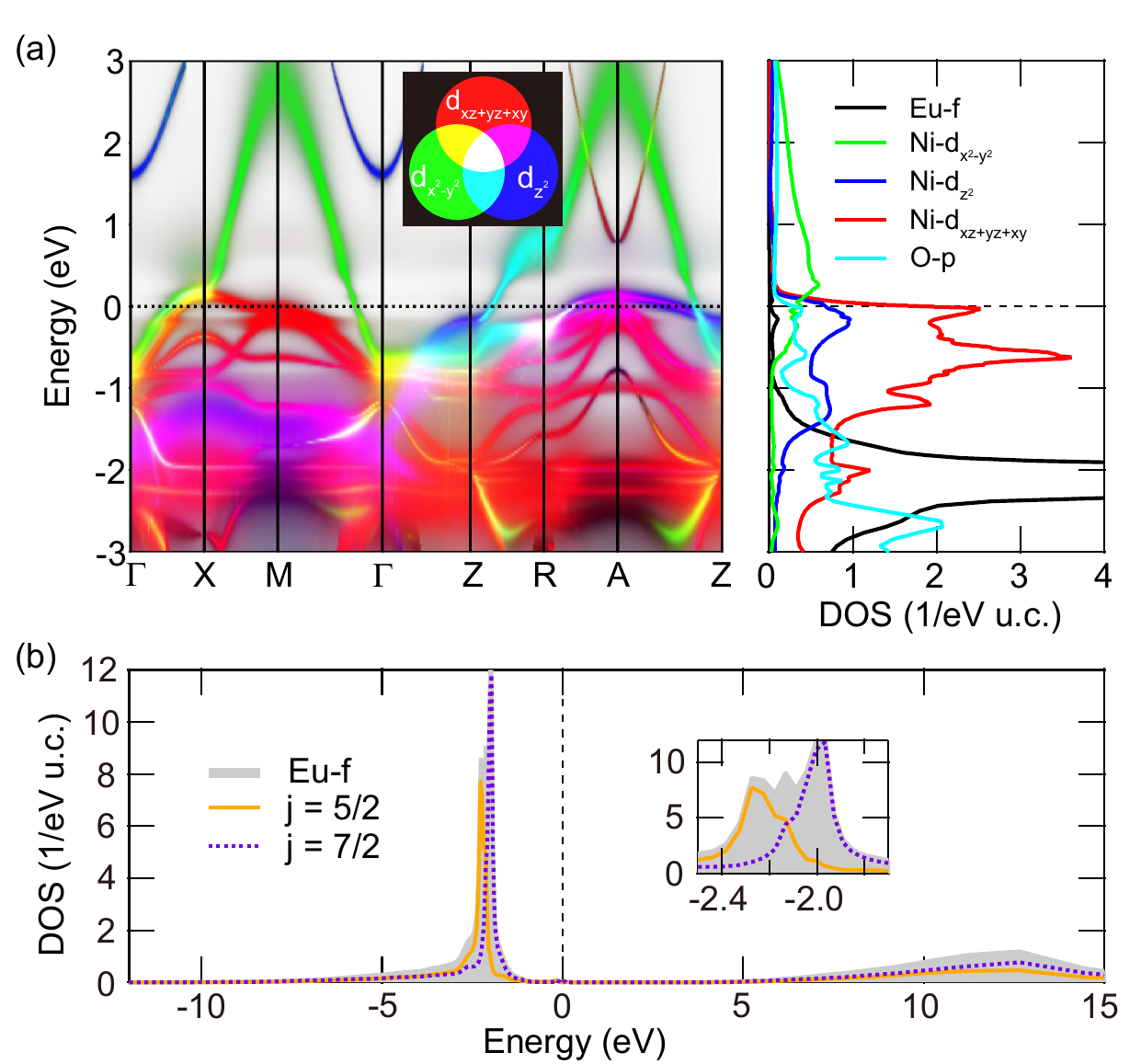}
        \caption{(a) Orbital-projected spectral function of EuNiO$_2$ from DFT+DMFT calculations at $T=300$\,K, with $U^{\rm Ni}=4$ eV, $J_{\rm H}^{\rm Ni}=0.7$ eV for Ni-$d$ orbitals and $U=8$ eV, $J_{\rm H}=1$ eV for Eu-$4f$ orbitals. Green, blue, and red colors denote the $d_{x^2-y^2}$, $d_{z^2}$, and $d_{xz}+d_{yz}+d_{xy}$ contributions of Ni ions, respectively. The right panel shows the orbital-resolved density of states (DOS), obtained by momentum integration, for the Ni-$3d$, O-$p$, and Eu-$4f$ orbitals. (b) Comparison of the partial densities of states of the $j=5/2$ and $7/2$ orbitals of Eu ions, showing evident lower and upper Hubbard bands well below and above the Fermi energy. The inset shows a magnified view of the spin-orbit split lower Hubbard band.}\label{fig2}
    \end{center}
\end{figure}

Figure \ref{fig2} presents the DFT+DMFT spectral function of EuNiO$_2$ at 300 K, computed with $U=$ 8 eV and $J_{\rm H}=$ 1 eV on Eu-4$f$ orbitals. As shown in Fig.~\ref{fig2}(b), the 4$f$ electrons are now strongly localized, giving rise to the lower Hubbard bands centered around -2.2 eV and the upper Hubbard bands above 7 eV. The integrated Eu-4$f$ occupation is approximately 7, indicating a divalent (Eu$^{2+}$) state. The $j=5/2$ and $7/2$ multiplets are both half filled and contribute to the Hubbard bands, as shown in the enlarged plot around -2.2 eV (inset). Correspondingly, the Ni-$d$ occupation decreases relative to the GGA prediction, indicating that the system is in a highly overdoped regime. The $t_{2g}$ and $d_{z^2}$ bands of Ni ions all shift upward and contribute a substantial density of states at $E_{\rm F}$, as evidenced in the right panel of Fig.~\ref{fig2}(a), whereas the $\beta'$ and $\beta$ Fermi pockets are significantly suppressed. The partially occupied Ni-$d$ bands remain highly itinerant with negligible mass renormalization.

\begin{figure}[tb]
    \centering
        \includegraphics[width=0.48\textwidth]{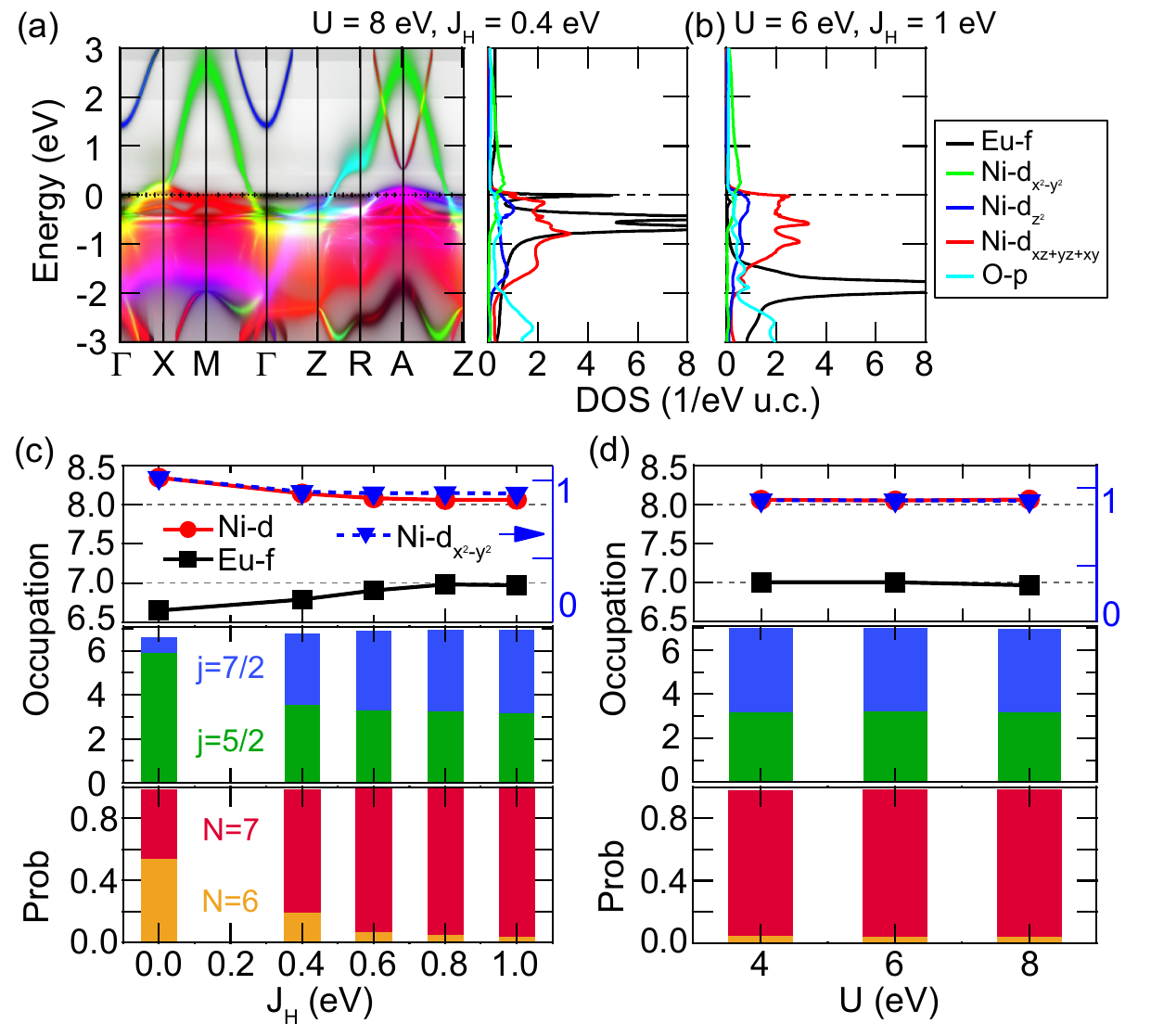}
        \caption{(a) Orbital-projected spectral function (left) and partial densities of states (right) of EuNiO$_2$ from DFT+DMFT calculations at 300\,K, with $U^{\rm Ni}=4$ eV, $J_{\rm H}^{\rm Ni}=0.7$ eV for Ni-$d$ orbitals and $U=8$ eV, $J_{\rm H}=0.4$ eV for Eu-$4f$ orbitals. (b) Partial densities of states at $U=6$ eV, $J_{\rm H}=1$ eV for comparison. (c),(d) Evolution of the electron occupation on different orbitals and the probability of $N=6$ and 7 configuration of Eu-$4f$ as functions of $J_{\rm H}$ at fixed $U=8$ eV and as functions of $U$ at fixed $J_{\rm H}=1$ eV, extracted from the DMFT impurity solver.}\label{fig3}
\end{figure}

To obtain the experimental Eu valence and clarify how it is changed by electronic correlations, we investigate the spectral function and orbital occupancies of Eu-4$f$ and Ni-3$d$ orbitals as functions of $U$ and $J_{\rm H}$ on the $4f$ shell. The results are summarized in Fig.~\ref{fig3}. No significant changes are found as $U$ varies from 8 to 4 eV for $J_{\rm H}=1$ eV as shown in Fig.~\ref{fig3}(d), indicating that the valence change is primarily driven by Hund's coupling rather than the Coulomb repulsion. A direct comparison between Figs.~\ref{fig2}(a) and \ref{fig3}(b) further confirms that reducing $U$ from 8 to 6 eV leaves the calculated densities of states (DOS) essentially unchanged. We therefore focus on the effect of $J_{\rm H}$ in what follows. 

As shown in Fig.~\ref{fig3}(c), DFT+DMFT calculations for $J_{\rm H}=0$ yield an Eu-4$f$ occupation of approximately 6.6. As $J_{\rm H}$ increases, the number is significantly enhanced and approaches 7.0 for $J_{\rm H} \geq 0.8$ eV, causing a dramatic change in the Eu-$4f$ configuration. At $J_{\rm H} = 0$, the system predominantly adopts a nonmagnetic configuration ($J=L-S=0$) with fully-filled $j=5/2$ multiplet and marginally occupied $j=7/2$ shell as predicted by pure DFT calculations. As the Hund's coupling is switched on, near half the electrons are redistributed into the $j=7/2$ orbitals, together with a sharp increase of the probability of the $N=7$ configuration. For $J_{\rm H}\gtrsim 0.8$\,eV, the $N=7$ configuration becomes dominant, with both the $j=5/2$ and 7/2 multiplets exhibiting half-filled occupations, signaling a divalent Eu state ($4f^7$).

\begin{figure}
    \begin{center}
        \includegraphics[width=0.48\textwidth]{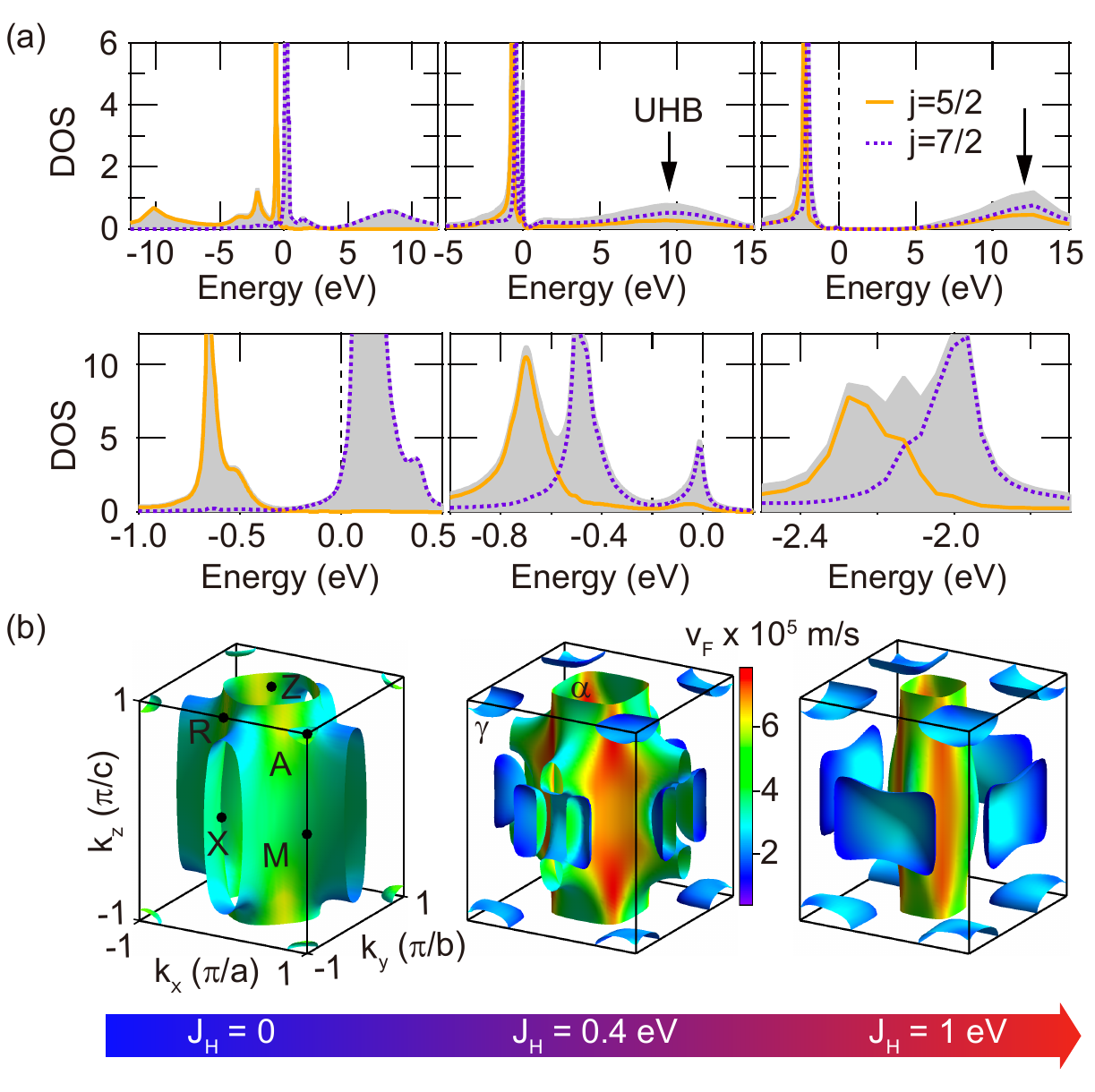}
        \caption{(a) Comparison of the $j=5/2$ and 7/2 densities of states of Eu ions over a wide energy range and a magnified window near $E_{\rm F}$ for $J_{\rm H}=0$, 0.4, and 1 eV for Eu-$4f$ orbitals. Black arrows mark the $4f$ upper Hubbard band pushed upward to high energy by the local Coulomb repulsion. (b) Evolution of the Fermi surface topology as $J_{\rm H}$ increases from 0 to 0.4 to 1 eV. The colors indicate the Fermi velocity.}\label{fig4}
    \end{center}
\end{figure}

More insights on the Eu ionic states can be obtained by analyzing the $j=5/2$ and 7/2 partial DOS plotted in Fig.~\ref{fig4}(a). In the $J_{\rm H}=0$ limit, the $j=5/2$ spectra lie almost entirely below the Fermi energy $E_{\rm F}$, while the $j=7/2$ spectra lie mostly above, with a small tail crossing the Fermi energy. The Eu-$4f$ state is therefore nonmagnetic and its low-energy property is governed by weakly correlated $j=7/2$ electrons. By contrast, for large $J_{\rm H}\gtrsim 0.8$ eV, both $j=5/2$ and 7/2 spectra show evident lower and upper Hubbard bands, with no evident quasiparticle peaks around the Fermi energy, indicating that both orbitals are in a half-filled Mott regime and the divalent Eu ions exhibit a large magnetic moment. In between, as $J_{\rm H}$ increases to a moderate value, the electrons transfer gradually from the $j=5/2$ orbitals to $j=7/2$ orbitals. At $J_{\rm H}=0.4$ eV, for example, the $j=5/2$ orbitals are half filled and located in a Mott regime, showing two well-separated Hubbard peaks below and above the Fermi energy, while the $j=7/2$ spectra exhibit a three-peak structure, with an additional sharp quasiparticle peak at the Fermi energy, indicating that the $j=7/2$ orbitals are close to half filling and located in a hole-doped Mott regime. Overall, increasing the Hund's coupling drives the Eu-$4f$ electrons toward a more spin-polarized configuration and enhances their magnetic character, while simultaneously redistributing spectral weight between the $j=5/2$ and $j=7/2$ manifolds. The Eu-$4f$ configuration consequently evolves from a predominantly filled $j=5/2$ manifold, characteristic of an approximately Eu$^{3+}$-like $4f^6$ configuration, toward half occupancy of both manifolds, approaching the Eu$^{2+}$ $4f^7$ configuration. The intermediate mixed-valence state exhibits orbital-selective Mott physics, characterized by coexisting $j=7/2$ quasiparticles and an enhanced magnetic moment desired in experiment.

Such local-itinerant duality of Eu-$4f$ electrons has important consequences on the fundamental physics of Eu-substituted infinite-layer nickelate superconductors. First, the Eu-$4f$ quasiparticle bands near the Fermi energy may hybridize with Ni-$3d$ orbitals and change the low-energy electronic structures. As an example, Fig.~\ref{fig4}(b) plots the Fermi surface evolution with $J_{\rm H}$. For $J_{\rm H}=0$, the Fermi surface topology is similar to other 112 compounds, consisting of a large cylindrical $\alpha$ sheet associated with the Ni-$d_{x^2-y^2}$ band and small hole pockets (interstitial $s$) centered at A, and the $\alpha$ Fermi surface exhibits electron-like character at $k_z=\pi$ and hole-like character at $k_z=0$. The hybridization with $4f$ quasiparticles at intermediate $J_{\rm H}$ marks a true difference of Eu-substituted compounds, giving rise to band bending and modified dispersions near the Fermi energy, as shown in Fig.~\ref{fig3}(a), reminiscent of heavy-fermion systems \cite{Shim2007S}, and Fermi-surface reconstruction shown in Fig.~\ref{fig4}(b). These produce non-negligible $4f$ contributions to the low-energy properties beyond the simple charge-reservoir picture of other rare-earth ions, which may be an important ingredient for eventually understanding the elevated $T_c$ observed in Eu-substituted 112 compounds. As $J_{\rm H}$ further increases and the Eu valence changes, the Ni-$3d$ (mainly $d_{x^2-y^2}$) orbitals become heavily hole-doped. The $\alpha$ Fermi surface then shrinks and undergoes a Lifshitz transition driven by the localization of Eu-$4f$ electrons. Second, the large magnetic moment of $4f$ orbitals may induce Kondo-type scattering of Ni-$d_{x^2-y^2}$ electrons as well as the long-range Ruderman-Kittel-Kasuya-Yosida (RKKY) interaction. While the former may typically affect the transport properties and lead to reentrant superconductivity as in (La,Ce)Al$_2$ \cite{Maple1972SSC}, the latter may be responsible for the ferromagnetic signals observed at low temperatures. The ferromagnetism may be further promoted by the Hund's coupling between $j=5/2$ and $7/2$ orbitals through a double-exchange mechanism. Such dual character of the mixed-valence Eu ions is essential for explaining the peculiar properties of Eu-substituted infinite-layer nickelate superconductors and should be taken into consideration in constructing their effective low-energy models.

To summarize, we have performed comprehensive DFT+DMFT calculations to investigate the spin and valence state of Eu ions in 112-type nickelates. The Hund's rule coupling is found to play a key role in polarizing the $j=5/2$ manifold and inducing electron transfer to the $j=7/2$ orbitals, thereby driving the Eu ions into a mixed-valence state at moderate $J_{\rm H}$. In this intermediate regime, the Eu ion is located in a hole-doped orbital-selective Mott regime, and featured with a large magnetic moment from nearly half-filled $4f$ orbitals and itinerant $j=7/2$ quasiparticles. While the Eu-$4f$ quasiparticles may hybridize with the Ni-$3d$ electrons and contribute to the superconductivity, its large magnetic moment supplies additional magnetic correlations that may play a key role in its magnetic responses and superconducting reentrance. Our work highlights the potential importance of Eu-$4f$ electrons and provides a theoretical framework for understanding the interplay of valence, spin, and electronic properties in Eu-substituted infinite-layer nickelate superconductors.

This work was supported by the High-Level Talent Research Start-Up Project Funding of Henan Academy of Sciences (Projects No. 20251827011, No. 242027151, and No. 241827010), the National Natural Science Foundation of China (Grant No. 12474136), and the National Key Research and Development Program of China (Grant No. 2024YFA1408602). Numerical computations were performed at the Hefei Advanced Computing Center.


\begin{thebibliography}{20}

\bibitem{Li2019N}	D. Li, K. Lee, B. Y. Wang, M. Osada, S. Crossley, H. R. Lee, Y. Cui, Y. Hikita, and H. Y. Hwang, Superconductivity in an infinite-layer nickelate, Nature \textbf{572}, 624--627 (2019).

\bibitem{Li2020PRL} D. Li, B. Y. Wang, K. Lee, S. P. Harvey, M. Osada, B. H. Goodge, L. F. Kourkoutis, and H. Y. Hwang, Superconducting dome in Nd$_{1-x}$Sr$_x$NiO$_2$ infinite layer films, Phys. Rev. Lett. \textbf{125}, 027001 (2020).

\bibitem{Zeng2022SA} S. Zeng et al., Superconductivity in infinite-layer nickelate La$_{1-x}$Ca$_{x}$NiO$_2$ thin films, Sci. Adv. \textbf{8}, eabl9927 (2022).

\bibitem{Osada2020NL} M. Osada, B. Y. Wang, B. H. Goodge, K. Lee, H. Yoon, K. Sakuma, D. Li, M. Miura, L. F. Kourkoutis, and H. Y. Hwang, A superconducting praseodymium nickelate with infinite layer structure, Nano Lett. \textbf{20}, 5735 (2020).

\bibitem{Osada2020PRM} M. Osada, B. Y. Wang, K. Lee, D. Li, and H. Y. Hwang, Phase diagram of infinite layer praseodymium nickelate Pr$_{1-x}$Sr$_{x}$NiO$_2$ thin films, Phys. Rev. Mater. \textbf{4}, 121801 (2020).

\bibitem{Wei2023SA} W. Wei, D. Vu, Z. Zhang, F. J. Walker, and C. H. Ahn, Superconducting Nd$_{1-x}$Eu$_x$NiO$_2$ thin films using in situ synthesis, Sci. Adv. \textbf{9}, eadh3327 (2023).

\bibitem{Parzyck2025PRX} C. T. Parzyck et al., Superconductivity in the parent infinite-layer nickelate ${\mathrm{NdNiO}}_{2}$, Phys. Rev. X \textbf{15}, 021048 (2025).

\bibitem{Karp2020PRX} J. Karp, A. S. Botana, M. R. Norman, H. Park, M. Zingl, and A. Millis, Many-body electronic structure of {NdNiO$_{2}$} and {CaCuO$_{2}$}, Phys. Rev. X \textbf{10}, 021061 (2020).

\bibitem{Jiang2019PRB} P. Jiang, L. Si, Z. Liao, and Z. Zhong, Electronic structure of rare-earth infinite-layer $R$NiO$_2$ ($R$=La,Nd), Phys. Rev. B \textbf{100}, 201106 (2019).

\bibitem{Hepting2020NM} M. Hepting et al., Electronic structure of the parent compound of superconducting infinite-layer nickelates, Nat. Mater. \textbf{19}, 381 (2020).

\bibitem{Been2021PRX} E. Been, W.-S. Lee, H. Y. Hwang, Y. Cui, J. Zaanen, T. Devereaux, B. Moritz, and C. Jia, Electronic structure trends across the rare-earth series in superconducting infinite-layer nickelates, Phys. Rev. X \textbf{11}, 011050 (2021).

\bibitem{Yang2022FP} Y.-F. Yang and G.-M. Zhang, Self-doping and the Mott-Kondo scenario for infinite-layer nickelate superconductors, Front. Phys. \textbf{9}, 801236 (2022).

\bibitem{Gu2022I} Q. Gu and H.-H. Wen, Superconductivity in nickel-based 112 systems, The Innovation \textbf{3}, 100202 (2022).

\bibitem{Zhang2020PRBa} G.-M. Zhang, Y.-F. Yang, and F.-C. Zhang, Self-doped Mott insulator for parent compounds of nickelate superconductors, Phys. Rev. B \textbf{101}, 020501 (2020).

\bibitem{Kitatani2020NQM} M. Kitatani, L. Si, O. Janson, R. Arita, Z. Zhong, and K. Held, Nickelate superconductors---a renaissance of the one-band Hubbard model, npj Quantum Mater. \textbf{5}, 59 (2020).

\bibitem{Cui2021CPL} Y. Cui, C. Li, Q. Li, X. Zhu, Z. Hu, Y.-F. Yang, J. Zhang, R. Yu, H.-H. Wen, and W. Yu, NMR evidence of antiferromagnetic spin fluctuations in Nd$_{0.85}$Sr$_{0.15}$NiO$_2$, Chin. Phys. Lett. \textbf{38}, 067401 (2021).

\bibitem{Wu2020PRB} X. Wu, D. Di Sante, T. Schwemmer, W. Hanke, H. Y. Hwang, S. Raghu, and R. Thomale, Robust $d_{x^2-y^2}$-wave superconductivity of infinite-layer nickelates, Phys. Rev. B \textbf{101}, 060504 (2020).

\bibitem{Wang2020PRBa} Z. Wang, G.-M. Zhang, Y.-F. Yang, and F.-C. Zhang, Distinct pairing symmetries of superconductivity in infinite-layer nickelates, Phys. Rev. B \textbf{102}, 220501 (2020).

\bibitem{Chen2023NC} H. Chen, Y.-F. Yang, G.-M. Zhang, and H. Liu, An electronic origin of charge order in infinite-layer nickelates, Nat. Commun. \textbf{14}, 5477 (2023).

\bibitem{Gu2020CP} Y. Gu, S. Zhu, X. Wang, J. Hu, and H. Chen, A substantial hybridization between correlated Ni-d orbital and itinerant electrons in infinite-layer nickelates, Commun. Phys. \textbf{3}, 84 (2020).

\bibitem{Nomura2019PRB} Y. Nomura, M. Hirayama, T. Tadano, Y. Yoshimoto, K. Nakamura, and R. Arita, Formation of a two-dimensional single-component correlated electron system and band engineering in the nickelate superconductor ${\mathrm{NdNiO}}_{2}$, Phys. Rev. B \textbf{100}, 205138 (2019).

\bibitem{Gu2020NC} Q. Gu et al., Single particle tunneling spectrum of superconducting Nd$_{1-x}$Sr$_x$NiO$_2$ thin films, Nat. Commun. \textbf{11}, 6027 (2020).

\bibitem{Tam2022NM} C. C. Tam et al., Charge density waves in infinite-layer NdNiO$_2$ nickelates, Nat. Mater. \textbf{21}, 1116 (2022).

\bibitem{Sun2025SA} W. Sun et al., Electronic structure of superconducting infinite-layer lanthanum nickelates, Sci. Adv. \textbf{11}, eadr5116 (2025).

\bibitem{Li2025PRL} C. Li et al., Observation of electridelike $s$ states coexisting with correlated $d$ electrons in  NdNiO$_2$, Phys. Rev. Lett. \textbf{135}, 116501 (2025).

\bibitem{Yang2026NC} M. Yang et al., Enhanced superconductivity and mixed-dimensional behaviour in infinite-layer samarium nickelate thin films, Nat. Commun. \textbf{17}, 2761 (2026).

\bibitem{Chow2025N} S. L. E. Chow, Z. Luo, and A. Ariando, Bulk superconductivity near 40 K in hole-doped SmNiO$_2$ at ambient pressure, Nature \textbf{642}, 58--63 (2025).

\bibitem{Han2026NSR} H. Han et al., A chemical avenue to manipulate field-reentrant superconducting competition in infinite-layer nickelates, Natl. Sci. Rev. nwag457 (2026).

\bibitem{Yang2026N} M. Yang et al., Field re-entrant superconductivity in Eu-doped infinite-layer nickelates, Nature \textbf{653}, 1052--1059 (2026).

\bibitem{Varbaro2026NC} L. Varbaro et al., Paramagnetically driven superconducting re-entrance in Eu-doped infinite layer nickelates, Nat. Commun. \textbf{17}, 7738 (2026).

\bibitem{Vu2026NC} D. Vu et al., Re-entrant unconventional superconductivity induced by rare-earth substitution in Nd$_{1-x}$Eu$_x$NiO$_2$ thin films, Nat. Commun. \textbf{17}, 3480 (2026).

\bibitem{Yang2026} W. Yang et al., Superconducting dome and field-enhanced superconductivity of PLD synthesized Nd$_{1-x}$Eu$_x$NiO$_2$ thin films, arXiv:2607.10332.

\bibitem{Grutter2025} A. J. Grutter et al., Net magnetization and inhomogeneous magnetic order in a high-$T_c$ nickelate superconductor, arXiv:2512.18005.

\bibitem{Blaha2014} P. Blaha, K. Schwarz, G. K. H. Madsen, D. Kvasnicka, and J. Luitz, WIEN2k: An Augmented Plane Wave + Local Orbitals Program for Calculating Crystal Properties (Vienna University of Technology, Austria, 2014).

\bibitem{Blaha2020JCP} P. Blaha, K. Schwarz, F. Tran, R. Laskowski, G. K. H. Madsen, and L. D. Marks, WIEN2k: An APW+lo program for calculating the properties of solids, J. Chem. Phys. \textbf{152}, 074101 (2020).

\bibitem{Perdew1996PRL} J. P. Perdew, K. Burke, and M. Ernzerhof, Generalized gradient approximation made simple, Phys. Rev. Lett. \textbf{77}, 3865 (1996).


\bibitem{Riley2018NC} J. M. Riley et al., Crossover from lattice to plasmonic polarons of a spin-polarised electron gas in ferromagnetic EuO, Nat. Commun. \textbf{9}, 2305 (2018).

\bibitem{Lazewski2021IC} J. \L a\.zewski, M. Sternik, P. T. Jochym, J. Kalt, S. Stankov, A. I. Chumakov, J. G\"ottlicher, R. R\"uffer, T. Baumbach, and P. Piekarz, Lattice dynamics and structural phase transitions in Eu$_2$O$_3$, Inorg. Chem. \textbf{60}, 9571 (2021).

\bibitem{Haule2007PRB} K. Haule, Quantum Monte Carlo impurity solver for cluster dynamical mean-field theory and electronic structure calculations with adjustable cluster base, Phys. Rev. B \textbf{75}, 155113 (2007).

\bibitem{Zeer2025NCM} M. Zeer, D. Go, M. Kl\"aui, W. Wulfhekel, S. Bl\"ugel, and Y. Mokrousov, Orbital torques and orbital pumping in two-dimensional rare-earth dichalcogenides, npj Comput. Mater. \textbf{11}, 305 (2025).

\bibitem{Jarrell1996PR} M. Jarrell and J. E. Gubernatis, Bayesian inference and the analytic continuation of imaginary-time quantum Monte Carlo data, Phys. Rep. \textbf{269}, 133 (1996).

\bibitem{Haule2015PRL} K. Haule, Exact double counting in combining the dynamical mean field theory and the density functional theory, Phys. Rev. Lett. \textbf{115}, 196403 (2015).

\bibitem{Shim2007S} J. H. Shim, K. Haule, and G. Kotliar, Modeling the localized-to-itinerant electronic transition in the heavy fermion system {CeIrIn$_5$}, Science \textbf{318}, 1615 (2007).

\bibitem{Maple1972SSC} M. B. Maple, W. A. Fertig, A. C. Mota, L. E. DeLong, D. Wohlleben, and R. Fitzgerald, The re-entrant superconducting-normal phase boundary of the Kondo system (La,Ce)Al$_2$, Solid State Commun. \textbf{11}, 829 (1972).

\end{thebibliography}
\end{document}